\documentclass[aps,prl,twocolumn,superscriptaddress,nofootinbib,longbibliography]{revtex4-2}

\usepackage[utf8]{inputenc}
\usepackage[T1]{fontenc}
\usepackage{lmodern}
\usepackage{amsmath,amssymb,bm}
\usepackage{graphicx}
\usepackage{xcolor}
\usepackage[hidelinks]{hyperref}
\usepackage{microtype}
\providecommand{\JournalTitle}[1]{\emph{#1}}
\providecommand{\enquote}[1]{``#1''}

\begin{document}

\title{Geometric phase of arbitrary Mueller evolutions and its two-level quantum analogue}

\author{Jos\'e J. Gil}
\email{ppgil@unizar.es}
\affiliation{Independent Researcher, 29690 Casares Costa, Spain}

\begin{abstract}
We identify, for a general physically realizable Mueller transformation, the only intrinsic geometric-phase structure that can be assigned to it in an invariant manner: the retarding part of the characteristic pure component selected by the characteristic decomposition, which defines a canonical holonomic content. A Mueller matrix does not, in general, determine a unique observed interferometric (Pancharatnam) geometric phase, since the latter depends on the specific physical realization of the transformation and on the interferometric readout. The remaining characteristic layers may modify the measured complex visibility, and even its observed argument through convex averaging, but they do not define a unique geometric holonomy of their own. We further establish the quantum analogue for open two-level dynamics within the Choi representation.
\end{abstract}

\maketitle

Geometric phase is a manifestation of holonomy associated with parallel transport and plays a central role in systems ranging from polarized light to quantum two-level dynamics. Building on the identification of the antisymmetric Mueller generator as the universal algebraic kernel of geometric phase for ideal retarders and unitary two-level evolutions~\cite{GilJOSAA2026AMG}, we extend that viewpoint to arbitrary (even strongly depolarizing) Mueller maps by isolating their unique pure characteristic core. Here we use “geometric phase” in the interferometric (Pancharatnam) sense, i.e., the argument of the complex visibility, with dynamical contributions removed or calibrated when appropriate~\cite{Pancharatnam1956,Berry1984,SamuelBhandari1988,Sjoqvist2000}. For ideal unitary evolutions of two-level systems, the geometric phase is fixed by the rotational (Hamiltonian) generator. Beyond that deterministic/unitary setting, the present formulation provides, for general physically realizable Mueller transformations, a practical recipe to extract from a measured Mueller matrix both (i) the canonical geometric-phase carrier, through the antisymmetric Mueller generator of the characteristic pure core, and (ii) the first index of polarimetric purity $P_1$, which gives the coherent weight of that core and sets the corresponding interferometric visibility scale. This can be useful for interpreting geometric-phase interferometry in the presence of depolarization, for robust design/tolerance analysis of polarization elements, and for a geometric diagnosis of qubit channels via their characteristic unitary core. The prescription can be tested in standard interferometric setups by implementing controlled depolarizing layers (e.g., retarder ensembles or polarization scramblers) in one arm.

Accordingly, the aim here is to identify the intrinsic geometric content that can still be assigned to a general physically realizable Mueller transformation. While such a matrix does not, in general, determine a unique interferometric geometric phase---because the latter depends on the specific physical realization and on the interferometric readout---the characteristic decomposition still singles out a unique coherent Mueller--Jones component whose retarding part defines a canonical holonomic content. This is the only intrinsic geometric-phase structure that can be associated with the transformation in an invariant manner.

In realistic scenarios, evolutions are frequently non-unitary. In classical optics, physically realizable Mueller transformations generally involve depolarization, whereas in quantum mechanics open-system dynamics is described by completely positive trace-preserving (CPTP) maps that induce decoherence and mixing. In such cases, it is not \emph{a priori} clear which part of the transformation genuinely carries geometric holonomy and which part merely reduces coherence.

From an experimental viewpoint, Mueller polarimetry provides a physically realizable mapping on Stokes space through a measured Mueller matrix $\mathbf{M}$, i.e., an input--output ``jump'' that may include substantial depolarization. A natural question is then whether (and how) an interferometric geometric phase can be assigned to such a transformation in a unique and physically meaningful way. This is nontrivial because a Mueller matrix carries no global phase information, and because depolarizing layers may contain internal rotational content at the level of their pure constituents. Existing treatments typically address purely deterministic (retarder-like) evolutions or mixed-state phases in quantum interferometry, but they do not provide a direct, invariant prescription to identify the genuinely geometric contribution from a general measured Mueller matrix.

Our approach starts by isolating, in an invariant way, the maximally weighted coherent (pure) core of a measured Mueller matrix through the \emph{characteristic decomposition}~\cite{Gil2007PolarimetricCO,GilStructure2016}. Let $\mathbf{M}$ be a physically realizable Mueller matrix with mean-intensity coefficient $m_{00}$. We define the normalized Mueller matrix
\begin{equation}
\hat{\mathbf{M}}=\frac{1}{m_{00}}\,\mathbf{M},
\end{equation}
since an overall gain does not affect interferometric phases. The characteristic decomposition of $\hat{\mathbf{M}}$ reads~\cite{Gil2007PolarimetricCO,GilStructure2016}
\begin{equation}
\hat{\mathbf{M}}
= P_1\hat{\mathbf{M}}_{J} + (P_2-P_1)\hat{\mathbf{M}}_{(2)} + (P_3-P_2)\hat{\mathbf{M}}_{(3)} + (1-P_3)\hat{\mathbf{M}}_{\Delta0},
\label{eq:char-decomp-full}
\end{equation}
where $0\le P_1\le P_2\le P_3\le 1$ are the indices of polarimetric purity (IPP), $\hat{\mathbf{M}}_{J}$ is the pure (Mueller--Jones) characteristic core, and $\hat{\mathbf{M}}_{\Delta0}$ is the maximally mixed (fully random) depolarizer. The remaining mixed structure can be grouped into a single \emph{discriminating} contribution, $\hat{\mathbf{M}}_{\mathrm{disc}}\propto (P_2-P_1)\hat{\mathbf{M}}_{(2)}+(P_3-P_2)\hat{\mathbf{M}}_{(3)}$, so that the characteristic decomposition can be viewed (conceptually) as the sum of three parts: pure core, discriminating, and maximally mixed.

The IPP are defined from the eigenvalues of the associated covariance (coherency) matrix $\mathbf{H}$~\cite{Cloude1986GroupTA,arnalmodelo1990,SanJoseGil2011}, ordered by convention as $\lambda_0\ge\lambda_1\ge\lambda_2\ge\lambda_3\ge0$,

\begin{equation}
P_1=\lambda_0-\lambda_1,\quad
P_2=\lambda_0+\lambda_1-2\lambda_2,\quad
P_3=\lambda_0+\lambda_1+\lambda_2-3\lambda_3.
\label{eq:IPP-def}
\end{equation}
With this ordering, the pure characteristic component $\hat{\mathbf{M}}_{J}$ coincides as a Mueller--Jones matrix with the first (dominant) spectral pure component generated by the eigenvector associated with $\lambda_0$. Importantly, however, its weight in the characteristic decomposition is $P_1$ rather than $\lambda_0$; the difference reflects the redistribution of spectral content into the canonical depolarizing layers (including the fully random contribution), as detailed in Ref.~\cite{GilOptimalFiltering}. Moreover, $\hat{\mathbf{M}}_{(2)}$ and $\hat{\mathbf{M}}_{(3)}$ are uniquely defined as equiprobable mixtures of the first two and first three spectral pure components, respectively (in the same ordered-eigenvalue convention). In this way, the geometric-phase problem for arbitrary depolarizing Mueller systems reduces to the rotational content of a single Mueller--Jones core extracted invariantly from the measurement, while the remaining layers quantify visibility loss.

In other words, the characteristic decomposition provides a canonical separation between a single coherent Mueller--Jones core and a residual depolarizing part. We next connect this structure with the interferometric definition of geometric phase.

For a Mueller transformation with nonunit coherent weight $(P_1<1)$ it is convenient to group the non-pure contributions into a single normalized physically realizable component,
\begin{equation}
\hat{\mathbf{M}}_{\mathrm{np}}=
\frac{(P_2-P_1)\hat{\mathbf{M}}_{(2)}+(P_3-P_2)\hat{\mathbf{M}}_{(3)}+(1-P_3)\hat{\mathbf{M}}_{\Delta0}}{1-P_1},
\end{equation}
so that
\begin{equation}
\hat{\mathbf{M}}=P_1\hat{\mathbf{M}}_{J}+(1-P_1)\hat{\mathbf{M}}_{\mathrm{np}}.
\label{eq:M-hat-compact}
\end{equation}
The scalar $P_1$ thus quantifies the maximally weighted coherent fraction associated with a single Jones realization.

The remaining characteristic layers $\hat{\mathbf{M}}_{(2)}$, $\hat{\mathbf{M}}_{(3)}$, and $\hat{\mathbf{M}}_{\Delta0}$ encode progressively higher degrees of polarimetric randomness. These layers do not correspond to a single coherent evolution and therefore do not define a unique geometric holonomy. In an interferometric implementation, they may modify the measured complex visibility, and even its observed argument through convex averaging of complex amplitudes, but such effects do not constitute an intrinsic geometric phase of the Mueller transformation itself~\cite{GilOptimalFiltering}.

We further emphasize that the use of the characteristic decomposition is not merely a convenient choice but a physically and geometrically privileged one. On the one hand, it provides a direct description of the structure of polarimetric randomness through the nested characteristic layers selected by the indices of polarimetric purity~\cite{GilOptimalFiltering}. On the other hand, the indices of polarimetric purity, being ordered linear functions of the normalized eigenvalues of the associated covariance matrix, define a depolarization space with maximal volume among linear metric constructions in the space of physically admissible Mueller states \cite{OssikovskiVizet2019} and therefore possess privileged discriminating power. These features make the characteristic decomposition especially suited for the present problem, where both physical interpretability and structural optimality are essential.

We adopt an interferometric definition of geometric phase applicable to both classical polarization transformations and quantum two-level dynamics. In a two-arm interferometric arrangement, one arm undergoes the evolution under study while the other provides a reference. The complex visibility $\mathcal{V}$ is the overlap between the reference and the evolved field (or state), and the geometric phase is identified with $\arg(\mathcal{V})$, after subtraction of any dynamical contribution when required. Loss of coherence reduces $|\mathcal{V}|$, whereas the geometric phase is encoded in the argument.

It is also important to stress a basic point: obviously, a Mueller matrix does not encode any global optical phase. Even for a pure Mueller--Jones system, the associated Jones matrix is defined only up to an arbitrary scalar phase factor. The geometric phase addressed here is not a global phase, but a holonomy associated with coherent parallel transport. In this sense, Mueller polarimetry provides the physically realizable mapping on Stokes space (a ``jump'' from input to output), while the geometric phase is accessed through a coherent Jones realization of the pure component. The characteristic decomposition provides, in an invariant manner, the unique pure core that admits such a coherent interpretation, and whose rotational content fixes the holonomy.

The key technical step is therefore to extract the rotational content from the \emph{pure characteristic component}. For a general depolarizing Mueller matrix, applying polar-type factorizations is not, in general, physically meaningful in Mueller algebra (factors may fail to be Mueller-realizable), and the singular case introduces additional subtleties~\cite{GilOssikovskiSanJose2016Singular}. In contrast, $\hat{\mathbf{M}}_{J}$ is pure and admits a Jones realization, so its $3\times3$ block $\mathbf{m}_{J}$ can be decomposed unambiguously into a retarder (rotation) and a symmetric positive factor~\cite{GilBernabeu1987Optik,LuChipman1996}; in the singular case the decomposition is obtained by continuity~\cite{GilOssikovskiSanJose2016Singular}. Specifically, the polar decomposition of $\mathbf{m}_{J}$ reads
\begin{equation}
\mathbf{m}_{J}=\mathbf{m}_{R}\mathbf{m}_{D},\qquad \mathbf{m}_{R}\in SO(3),\quad \mathbf{m}_{D}=\mathbf{m}_{D}^{\mathsf T}\succ0,
\label{eq:polar-pure}
\end{equation}
Here $\mathbf{m}_{R}$ is the adjoint $SO(3)$ representation of the SU(2) \emph{unitary} factor of the Jones matrix associated with $\hat{\mathbf{M}}_{J}$ (i.e., the retarding part). Thus the antisymmetric Mueller generator extracted below is, explicitly, the generator of that unitary rotation; the symmetric factor $\mathbf{S}$ (dichroic/boost content) is neutral with respect to the geometric phase~\cite{GilJOSAA2026AMG}.

We then define the \emph{antisymmetric Mueller generator} (AMG) as
\begin{equation}
\mathbf{G}_{\mathrm{a}}=\log(\mathbf{m}_{R}),
\label{eq:AMG}
\end{equation}
where the logarithm is taken in the principal branch of $\mathfrak{so}(3)$, i.e., the associated rotation angle satisfies $\theta\in[0,\pi]$. In the deterministic retarder limit (no diattenuation, $\mathbf{S}=\mathbf{I}_3$), this definition reduces to the antisymmetric adjoint generator identified previously as the universal kernel of Pancharatnam--Berry phase for ideal retarders and unitary qubit evolutions~\cite{GilJOSAA2026AMG} up to the usual $2\pi$ branch ambiguity. Because a rigid rotation of the Poincar\'e (Bloch) sphere is generated solely by an antisymmetric generator, the AMG isolates the only part of the transformation that can produce parallel-transport holonomy; symmetric diattenuation/boost factors and depolarizing layers contract or distort the sphere and therefore cannot define geometric holonomy by themselves. The AMG thus represents the unique rotational content of the pure characteristic core and is the sole contributor to geometric holonomy.

The role of the mixed characteristic layers is best understood at the interferometric level, where phase is defined for complex field amplitudes rather than for Stokes vectors. Accordingly, we temporarily work at the field-amplitude level and represent a fully polarized input state by a normalized Jones spinor $|\psi\rangle\in\mathbb{C}^2$ (Dirac notation is used here only for compactness, not as a quantum assumption). Its associated normalized Stokes four-vector is $\hat{\mathbf{s}}=(1,\mathbf{u})^{T}$, where the unit Poincar\'e (Bloch) vector $\mathbf{u}\in\mathbb{R}^3$ is defined by
\begin{equation}
\mathbf{u}=\langle\psi|\boldsymbol{\sigma}|\psi\rangle,\qquad 
\boldsymbol{\sigma}\equiv(\sigma_x,\sigma_y,\sigma_z),
\end{equation}
with $\sigma_x,\sigma_y,\sigma_z$ the Pauli matrices. Here we adopt the standard polarization-optics ordering, for which the third matrix (our $\sigma_z$) is the purely imaginary one and represents the circular/helicity Stokes component (it corresponds to $\sigma_y$ in the conventional quantum-mechanics labeling). Accordingly, $u_i=\langle\psi|\sigma_i|\psi\rangle$ and $|\mathbf{u}|=1$ for fully polarized light. For each Mueller--Jones constituent, let $J_k$ denote a Jones realization (defined up to an overall phase, which cancels in the Mueller description). In a standard two-beam interferometric readout, where the reference arm preserves the input polarization and the detected interference term corresponds to projection onto $|\psi\rangle$, the complex visibility contributed by an equiprobable mixture is the convex sum of complex overlaps, e.g.,
\begin{equation}
\mathcal{V}_{(2)}=\tfrac12\langle\psi|J_1|\psi\rangle+\tfrac12\langle\psi|J_2|\psi\rangle,
\end{equation}
and similarly for $\mathcal{V}_{(3)}$ with three terms. The observed phase is $\arg(\mathcal{V}_{(2)})$ or $\arg(\mathcal{V}_{(3)})$, i.e., the argument of a sum of complex numbers, and in general it does not coincide with the phase of any constituent term.

A concrete example illustrates this point sharply. Let $J_1=\exp(-i\phi\sigma_z/2)$ and $J_2=\exp(-i\phi\sigma_x/2)$ be two unitary retarders and take $|\psi\rangle=|0\rangle$. Then $\langle 0|J_1|0\rangle=e^{-i\phi/2}$ while $\langle 0|J_2|0\rangle=\cos(\phi/2)$, and therefore
\begin{equation}
\mathcal{V}_{(2)}=\tfrac12 e^{-i\phi/2}+\tfrac12\cos(\phi/2)
=\cos(\phi/2)-\tfrac{i}{2}\sin(\phi/2),
\end{equation}
so that
\begin{equation}
\arg(\mathcal{V}_{(2)})=-\arctan\!\left(\tfrac12\tan(\phi/2)\right),
\end{equation}
which differs from the constituent holonomies (e.g., $-\phi/2$ for $J_1$). This phase is a property of the incoherent mixture and can exhibit non-holonomic behavior (including loss of definition when $|\mathcal{V}_{(2)}|\to 0$). Accordingly, mixed characteristic layers can contain \emph{internal} rotational content at the level of their pure constituents, yet they do not generate geometric holonomy.

A useful descriptor of the mixed structure is the \emph{discriminating} component introduced in Ref.~\cite{GilSymmetry2020Asymmetry}. For an $n$-dimensional density matrix $\rho$ (or, in the Mueller context, the normalized covariance matrix associated with $\hat{\mathbf{M}}$), define
\begin{equation}
\rho_{\mathrm{disc}}=\rho-P_1\rho_{1}-(1-P_{n-1})\rho_{n},
\label{eq:discriminating}
\end{equation}
i.e., the remainder after subtracting the pure characteristic term and the maximally mixed contribution $\rho_{n}=\mathbb{I}/n$. Nonregularity is signaled when $\rho_{\mathrm{disc}}$ is not real, equivalently when $\Im(\rho_{\mathrm{disc}})\neq 0$, revealing intrinsic antisymmetric correlations in the mixed sector. Such correlations are properties of the state (or layer) itself and must be distinguished from the antisymmetric part of a dynamical generator; accordingly, they can affect fringe visibility (e.g., by producing cancellations in convex sums), but they do not contribute to the geometric holonomy fixed by the AMG of the characteristic core. These considerations lead to the central statement: in the characteristic decomposition~\eqref{eq:char-decomp-full}, the geometric phase associated with $\hat{\mathbf{M}}$ is determined exclusively by the AMG of the pure characteristic component $\hat{\mathbf{M}}_{J}$, while $P_1$ quantifies the coherent weight (visibility scaling) of this contribution. This is precisely the generalization of our retarder-based result~\cite{GilJOSAA2026AMG}.

A particularly instructive limiting case is $P_1=0$, where the characteristic pure component is absent. In this situation, no canonical coherent core exists, and therefore no intrinsic geometric holonomy can be assigned to the Mueller transformation. Although specific physical implementations may still exhibit a nonzero interferometric phase, such a phase cannot be attributed uniquely to the transformation itself, but rather to the particular realization and averaging procedure.

The same question arises in open two-level quantum dynamics, where a general channel mixes unitary transport with decohering layers; our formulation shows that geometric holonomy is likewise fixed by a coherent characteristic core.

We finally outline the corresponding two-level quantum formulation and emphasize the distinction between \emph{states} and \emph{channels}. A density operator $\rho\in\mathbb{C}^{4\times4}$ satisfies $\rho=\rho^{\dagger}$, $\rho\succeq 0$, and $\mathrm{Tr}\,\rho=1$. Applying a characteristic decomposition to such a state identifies a dominant pure state and mixed layers, but it does not imply unitarity, which is a property of dynamical maps.

A qubit channel $\mathcal{E}$ (CPTP map) is represented by its Choi matrix~\cite{Jamiolkowski1972,Choi1975}
\begin{equation}
J_{\mathcal{E}}=(\mathcal{E}\otimes \mathrm{id})(|\Omega\rangle\langle\Omega|),\qquad
|\Omega\rangle=\tfrac{1}{\sqrt{2}}\sum_{i=0}^{1}|i\rangle\otimes|i\rangle,
\end{equation}
and the normalized Choi state $\rho_{\mathcal{E}}=\tfrac12 J_{\mathcal{E}}$. Trace preservation is equivalent to
\begin{equation}
\mathrm{Tr}_{\mathrm{out}}(\rho_{\mathcal{E}})=\mathbb{I}_2/2.
\end{equation}
Applying the characteristic decomposition to $\rho_{\mathcal{E}}$ yields a dominant rank-one component $\rho_{1}$ and mixed layers. The pure term $\rho_{1}$ can be reshaped into a single operator $K$ (a Kraus representative), i.e., $\rho_{1}\propto |K\rangle\!\rangle\langle\!\langle K|$. Note that $K$ is defined up to the standard unitary freedom in Kraus representations, which does not affect the polar decomposition of $K$ nor the geometric content of its unitary factor. In general $\rho_{1}$ need not be trace preserving: trace preservation of the full channel is enforced by the accompanying mixed layers. If $\mathrm{Tr}_{\mathrm{out}}(\rho_{1})=\mathbb{I}_2/2$ (equivalently $K^{\dagger}K=\mathbb{I}_2$), then $K$ is unitary and defines a Hamiltonian characteristic core whose rotational generator fixes the geometric phase. Otherwise (as in amplitude-damping type channels) $K$ is nonunitary, and the coherent holonomy is associated with the unitary factor of the polar decomposition of $K$, while the remaining layers encode dissipation and reduce visibility. Thus, the same structural principle holds: geometric holonomy is fixed by the antisymmetric generator of the coherent core, whereas mixed layers primarily reduce coherence.

This canonical holonomic core should be distinguished from the total retardance content obtained through serial decompositions of a Mueller matrix, which may isolate entrance and exit retarders \cite{Ossikovski2009,GilSanJoseOssikovski2024} but do not, by themselves, single out a unique coherent contribution.

We have established a unified geometric viewpoint for classical polarization and qubit dynamics in the presence of depolarization or decoherence. A general Mueller transformation does not determine a unique observable geometric phase. What it determines, through its characteristic decomposition, is a unique coherent Mueller--Jones core whose retarding part defines the canonical holonomic content of the transformation. This exhausts the geometrical information that can be intrinsically associated with a Mueller matrix. Any additional phase effects observed in interferometric measurements arise from the specific physical realization and from convex averaging, rather than from an intrinsic geometric property of the transformation itself.

\section*{Declarations}
\noindent Conflict of interest. The author declares no conflicts of interest.\par
\noindent Data availability. No additional data were generated or analyzed in this study.


\begin{thebibliography}{10}

\bibitem{GilJOSAA2026AMG}
J.~J. Gil, \enquote{Antisymmetric {Mueller} generator as the universal origin of geometric phase in classical polarization and quantum two-level systems,} {\protect\JournalTitle{J. Opt. Soc. Am. A}} \textbf{43}, 507--515 (2026).

\bibitem{Pancharatnam1956}
S.~Pancharatnam, \enquote{Generalized theory of interference, and its applications. part {I}. coherent pencils,} {\protect\JournalTitle{Proc. Indian Acad. Sci. A}} \textbf{44}, 247--262 (1956).

\bibitem{Berry1984}
M.~V. Berry, \enquote{Quantal phase factors accompanying adiabatic changes,} {\protect\JournalTitle{Proc. R. Soc. Lond. A}} \textbf{392}, 45--57 (1984).

\bibitem{SamuelBhandari1988}
J.~Samuel and R.~Bhandari, \enquote{General setting for {B}erry's phase,} {\protect\JournalTitle{Phys. Rev. Lett.}} \textbf{60}, 2339--2342 (1988).

\bibitem{Sjoqvist2000}
E.~Sj{\"o}qvist, A.~K. Pati, A.~Ekert, \emph{et~al.}, \enquote{Geometric phases for mixed states in interferometry,} {\protect\JournalTitle{Phys. Rev. Lett.}} \textbf{85}, 2845--2849 (2000).

\bibitem{Gil2007PolarimetricCO}
J.~J. Gil, \enquote{Polarimetric characterization of light and media -- physical quantities involved in polarimetric phenomena,} {\protect\JournalTitle{Eur. Phys. J. Appl. Phys.}} \textbf{40}, 1--47 (2007).

\bibitem{GilStructure2016}
J.~J. Gil, \enquote{Structure of polarimetric purity of a {Mueller} matrix and sources of depolarization,} {\protect\JournalTitle{Opt. Commun.}} \textbf{368}, 165--173 (2016).

\bibitem{Cloude1986GroupTA}
S.~R. Cloude, \enquote{Group theory and polarisation algebra,} {\protect\JournalTitle{Optik}} \textbf{75}, 26--36 (1986).

\bibitem{arnalmodelo1990}
P.~M. Arnal, \enquote{Modelo matricial para el estudio de fen{\'o}menos de polarizaci{\'o}n de la luz,} Ph.D. thesis, Universidad de Zaragoza, Zaragoza, Spain (1990).

\bibitem{SanJoseGil2011}
I.~San~Jos{\'e} and J.~J. Gil, \enquote{Invariant indices of polarimetric purity. generalized indices of purity for $n\times n$ covariance matrices,} {\protect\JournalTitle{Opt. Commun.}} \textbf{284}, 38--47 (2011).

\bibitem{GilOptimalFiltering}
J.~J. Gil, \enquote{On optimal filtering of measured {Mueller} matrices,} {\protect\JournalTitle{Appl. Opt.}} \textbf{55}, 5449--5455 (2016).

\bibitem{OssikovskiVizet2019}
R.~Ossikovski and J.~Vizet, \enquote{Eigenvalue-based depolarization metric spaces for {Mueller} matrices,} {\protect\JournalTitle{J. Opt. Soc. Am. A}} \textbf{36}, 1173--1186 (2019).

\bibitem{GilOssikovskiSanJose2016Singular}
J.~J. Gil, R.~Ossikovski, and I.~San~Jos{\'e}, \enquote{Singular {Mueller} matrices,} {\protect\JournalTitle{J. Opt. Soc. Am. A}} \textbf{33}, 600--609 (2016).

\bibitem{GilBernabeu1987Optik}
J.~J. Gil and E.~Bernab{\'e}u, \enquote{Obtainment of the polarizing and retardation parameters of a non-depolarizing optical system from the polar decomposition of its {Mueller} matrix,} {\protect\JournalTitle{Optik}} \textbf{76}, 67--71 (1987).

\bibitem{LuChipman1996}
S.-Y. Lu and R.~A. Chipman, \enquote{Interpretation of {M}ueller matrices based on polar decomposition,} {\protect\JournalTitle{J. Opt. Soc. Am. A}} \textbf{13}, 1106--1113 (1996).

\bibitem{GilSymmetry2020Asymmetry}
J.~J. Gil, \enquote{Sources of asymmetry and the concept of nonregularity of n-dimensional density matrices,} {\protect\JournalTitle{Symmetry}} \textbf{12}, 1002 (2020).

\bibitem{Jamiolkowski1972}
A.~Jamio{\l}kowski, \enquote{Linear transformations which preserve trace and positive semidefiniteness of operators,} {\protect\JournalTitle{Rep. Math. Phys.}} \textbf{3}, 275--278 (1972).

\bibitem{Choi1975}
M.-D. Choi, \enquote{Completely positive linear maps on complex matrices,} {\protect\JournalTitle{Linear Algebra Appl.}} \textbf{10}, 285--290 (1975).

\bibitem{Ossikovski2009}
R.~Ossikovski, \enquote{Analysis of depolarizing {Mueller} matrices through a symmetric decomposition,} {\protect\JournalTitle{J. Opt. Soc. Am. A}} \textbf{26}, 1109--1118 (2009).

\bibitem{GilSanJoseOssikovski2024}
J.~J. Gil, I.~San~Jos{\'e}, and R.~Ossikovski, \enquote{Characterization of retardance of nondepolarizing and depolarizing media,} {\protect\JournalTitle{J. Opt. Soc. Am. A}} \textbf{41}, 1544--1553 (2024).

\end{thebibliography}
\end{document}